\documentclass[fleqn,usenatbib]{mnras}

\usepackage{newtxtext,newtxmath}

\usepackage[T1]{fontenc}

\DeclareRobustCommand{\VAN}[3]{#2}
\let\VANthebibliography\thebibliography
\def\thebibliography{\DeclareRobustCommand{\VAN}[3]{##3}\VANthebibliography}

\usepackage{graphicx}	
\usepackage{amsmath}	
\usepackage{soul}	
\usepackage{multirow}

\usepackage{newtxtext,newtxmath}

\makeatletter
\newcommand*{\rom}[1]{\expandafter\@slowromancap\romannumeral #1@}

\usepackage{xcolor,colortbl}
\definecolor{mygreen}{RGB}{0,153,76}

\title[Ionization rates and non-thermal emissions from starburst nuclei]{Cosmic-ray induced ionization rates and non-thermal emissions from nuclei of starburst galaxies}

\author[V. H. M. Phan et al.]{
Vo Hong Minh Phan,$^{1,2}$\thanks{vhmphan@obspm.fr} Enrico Peretti,$^{3,4}$\thanks{enrico.peretti.science@gmail.com} Pierre Cristofari,$^{5}$ Antoine Gusdorf\,$^{1,6}$ and Philipp Mertsch$^{2}$
\\
$^{1}$ Sorbonne Université, Observatoire de Paris, PSL Research University, LERMA, CNRS UMR 8112, 75005 Paris, France
\\
$^{2}$ Institute for Theoretical Particle Physics and Cosmology (TTK), RWTH Aachen University, 52056 Aachen, Germany\\
$^{3}$APC, Universit\'e Paris Diderot, CNRS/IN2P3, CEA/Irfu, Observatoire de Paris, Sorbonne Paris Cit\'e, France \\
$^{4}$ Niels Bohr International Academy, Niels Bohr Institute, University of Copenhagen, Blegdamsvej 17, DK-2100 Copenhagen, Denmark\\
$^{5}$ Observatoire de Paris, PSL Research University, LUTH, 5 Place J. Janssen, 92195 Meudon, France\\
$^{6}$ Laboratoire de Physique de l’ENS, ENS, Universit\'e PSL, CNRS, Sorbonne Universit\'e, Universit\'e de Paris, 75005 Paris, France
}

\date{Accepted XXX. Received YYY; in original form ZZZ}

\pubyear{2024}

\newcommand*\df{\mathop{}\!\mathrm{d}}
\newcommand*\mpr{\mathop{}\!m_\mathrm{p}}

\newcommand{\n}{\nonumber}

\defcitealias{behrens2022}{B22}
\defcitealias{holdship2022}{H22}

\begin{document}
\label{firstpage}
\pagerange{\pageref{firstpage}--\pageref{lastpage}}
\maketitle

\begin{abstract}
Cosmic rays are the only agent capable of ionizing the interior of dense molecular clouds and, thus, they are believed to play an essential role in determining the physical and chemical evolution of star-forming regions. In this work, we aim to study cosmic-ray induced ionization rates in starburst environments using non-thermal emissions of cosmic rays from starburst nuclei. To this end, we first revisit cosmic-ray models which could explain data of non-thermal emissions from radio to X-ray and gamma-ray from nuclei of three prototypical starburst galaxies NGC 253, M82, and Arp 220. These models are then applied to predict ionization rates in starburst environments which gives values around $10^{-14}$ s$^{-1}$. Such a high value of the ionization rate, which is 2 to 3 orders of magnitude higher than the typical values found in the Milky Way, is probably due to relatively high rates of supernova explosions occurring within the nuclei of these starburst galaxies. We also discuss in more details the case of NGC 253 where our predicted ionization rate is found to be, in most cases, a few times smaller than the values inferred from molecular line observations of clouds in the starburst nucleus. The general framework provided in this work illustrates how the use of non-thermal emission data could help to provide more insights into ionization rates or, more generally, cosmic-ray impact in starburst environments.    
\end{abstract}




\section{Introduction}

Starburst galaxies are galaxies with high star formation rates typically ranging from $10$ to $10^3$ times higher than that of the Milky Way \citep{gao2004}. 
The intense star-forming activity also results in a high rate of supernova explosions and, as massive stars and supernova remnants (SNRs) are commonly believed to be cosmic-ray sources \citep{strong2007,grenier2015,gabici2019,cristofari2021}, starburst galaxies are expected to be filled with cosmic rays (CRs). 
The connection between CRs and the star forming activity in starburst galaxies is further supported by tight correlations between the inferred star formation rate and the non-thermal luminosity, ascribed to CRs \citep[see e.g.][and references therein]{Lacki2010,Kornecki2020,Ajello2020,Kornecki2022}.
Of particular interest are the starburst nuclei \citep[SBNi, see e.g.][]{Westmoquette_2009} which have sizes of about a few hundred parsecs but within these small region exhibit rates of supernova explosions comparable to or even higher than that of the entire Milky Way. 
It is for this reason that SBNi are considered ideal laboratories to study CR impact on star-forming regions. 
In these environments, the density of gas, radiation and magnetic field are inferred to be at least $10^2$ times larger than the average interstellar medium (ISM) of the Milky Way.
This implies that most of the injected non-thermal particles are expected to lose their energy before being able to escape \citep{Yoast-Hull_2013,peretti2019}. 
Such a calorimetric behavior for CRs, together with the enhanced star formation rate inferred at redshift 1-4 \citep{Madau2014} suggested that starbursts could be an ideal source class to substantially contribute to the diffuse flux of high-energy gamma-rays and neutrinos \citep{Tamborra2014,Bechtol2017,Peretti2020,Ambrosone2021,Roth2021,Owen2022,Peretti2022,Condorelli2023}.

It has long been suggested that CRs can play an essential role in setting the chemistry and even dynamics of star-forming regions \citep{padovani2020,gabici2022}. This is because these particles could penetrate deep inside dense molecular clouds, where X-rays and UV photons cannot penetrate, to ionize the interior of these objects \citep[e.g.][]{padovani2009,ivlev2018,phan2018,owen2021}. In other words, CRs control the ionization rate which is a key parameter in regulating the abundances of different chemical species in molecular clouds. The ionization rate also determines the coupling between gas and magnetic fields which is of critical importance for the process of star formation \citep{krumholz2019,girichidis2020}. Thus, the impact of CRs on star-forming regions can be partially quantified using the cosmic-ray ionization rate. 

In fact, there are a few different variants for the definition of the ionization rate \citep{neufeld2017}. Here, we refer to the ionization rate as the production rate of H$^+_2$ per hydrogen molecule which will be denoted as $\zeta(\rm H_{2})$.
The observational determination of $\zeta ({\rm H}_2)$ in the interstellar medium is a longstanding problem. It is usually solved by identifying a set of species whose abundance or abundance ratio is sensitive to it. As such, the method is intrinsically subject to various limitations. First, from the observational point of view, a sufficient number of lines and an appropriate radiative transfer treatment must be implemented in order to infer reliable column densities. Second, the model that is used to predict abundances or abundance ratios must contain all the chemical pathways relevant to the destruction and formation of the observed species, and all the excitation mechanisms (shocks, energetic photons in particular ionizing UV ones, and CRs) relevant to the observed regions. Last limitation, identifying which species is tracing which mechanism requires the computing of large grids of models purposefully covering all input parameters. Within this framework, various species have been found to allow for the determination of the cosmic ray ionization state in various environments within the limits of carefully spelt-out assumptions. Several reviews focus on the determination of the ionization rate in the dense and diffuse medium, like \citet{indriolo2013}, \citet{neufeld2017}, and \citet{barger2020}. In quiescent and dense regions completely shielded from dissociating UV radiation, where all hydrogen is in molecules, \citet{guelin1977} and \citet{wootten1979} analytically showed that the DCO$^+$ to HCO$^+$ abundance ratio can be used to measure the ionization rate in steady-state conditions. Their result was somehow confirmed by the use of a more complex chemical code with the same assumptions and applied to then-state-of-the-art observations by \citet{caselli1998}. In the diffuse molecular medium with electron abundance between 10$^{-7}$ and 10$^{-2}$, H$_3^+$ has been extensively used to measure the ionization rate, based on a simplistic description of its chemistry (e.g. \citealt{mccall2003}, \citealt{indriolo2007}, \citealt{indriolo2012}). In fact, the modelling works of \citet{lepetit2004} mitigated the conclusions by highlighting that constraining the ionization rate can not be done independently from other species. In other words, they showed that self-consistent models should be used to reproduce the abundances of both H$_3^+$ and a number of atomic and molecular species in order to effectively provide a measure of the ionization rate. \citet{lepetit2016} then detailed all the dependences of the H$_3^+$ abundance to parameters of such self-consistent models and used them to measure the ionization rate in the central molecular zone, adding complementary constraints on hydride abundances. Comprehensive modelling was also the path chosen by \citet{neufeld2017} to provide constraints on the ionization rate in diffuse (both atomic and molecular) clouds of the Galactic disk. 

In the NGC253 starburst galaxy, \citet{holdship2022} (hereafter \citetalias{holdship2022}) used ALMA observations to constrain the ionization rate to between $10^{-14}$~s$^{-1}$ and $8\times10^{-13}$~s$^{-1}$, using rather strong assumptions. Their observations were performed at 1$\fs$6 angular resolution, that is about 30 pc spatial resolution, for which they used a \lq single point model’ with no photo-processes included. They ruled out the possibility that shocks or UV radiation could play a role in the heating or in the modelling of the column densities of SO and H$_3$O$^+$. Their justification for the shock process omission was that parametric, non-self-consistent shock models with high pre-shock density predicted values of the abundance ratio that they did not find in the observations. Their observed abundance ratio was obtained from the non-local thermodynamic equilibrium code RADEX modelling of 3 lines of H$_3$O$^+$ and 37 lines of SO \citep[see][for more details on RADEX]{vandertak2007}. For both species, collisions with other particles than H$_2$ were not considered, and for H$_3$O$^+$, the three transitions they used have the upper energy level between 79.5 and 169.1 K only, possibly biasing the results towards a hot component. Additionnally, \citet{behrens2022} (from here on referred to as \citetalias{behrens2022}) used HNC and HCN observations from the same dataset as \citetalias{holdship2022} to constrain the ionization rate to within the range from $10^{-13}$ s$^{-1}$ to $10^{-12}$ s$^{-1}$. They used the same kind of chemical and radiative transfer modelling as \citetalias{holdship2022}, also neglecting UV radiation and shock processes. Given all these assumptions, related to the observations, radiative transfer and physico-chemical modelling, plus additional ones discussed in these two articles, we consider these ionization rate data as indications rather than definitive values. 
These results are, however, encouraging as they point to high values of $\zeta(\rm H_{2})$ which are qualitatively expected given the high supernova rate in the SBN of NGC 253. 
The relatively large difference in results of these similar analyses calls for a different approach to determine the ionization rate in these complex environments. In this paper, we will discuss a rather different approach to predict ionization rates in SBNi which relies on non-thermal emissions from these objects. 

Many SBNi are bright gamma-ray sources both in the GeV and TeV energy range and some of them are also detected with X-ray telescopes \citep{acciari2009,acero2009,ackermann2012,fleischhack2015}. Several prototypical starburst galaxies, for example NGC 253, M82, or Arp 220, have been observed by both satellite and ground-based gamma-ray telescopes revealing their gamma-ray spectrum extending from a few hundred MeVs to about 10 TeV \citep[see e.g.][and references therein]{Ajello2020,tibaldo2021}. 
The gamma-ray emission is likely dominated by the decay of neutral pions produced in interactions between CR protons and interstellar gas in SBNi independent on the transport conditions \citep{peretti2019,krumholz2020}. 
This means that the gamma-ray spectrum could be employed to extract the CR proton spectrum within these SBNi. 
In addition, CR electrons in these systems can also induce detectable emissions in the range extending from radio (with frequency around 1 GHz) to X-ray (around 1 keV) via synchrotron radiation. Observations in the X-ray domain could, however, be contaminated by unresolved sources such as X-ray binaries \citep{wik2014,strickland2007,paggi2017} and part of the radio emissions can also come CR electrons confined in a larger halo surrounding the SBN \citep{Yoast-Hull_2013}.
Nevertheless, the radio and X-ray spectrum can be used as an upper limit to constrain the CR electron spectrum. The combination of these non-thermal emissions from radio to X-ray and gamma-ray provides a powerful tool to study CRs in starburst galaxies and, ultimately, allows us to quantify the impact of these particles in the star forming activity of these systems.

The paper will be structured as follows. In Section \ref{sec:transport}, we will introduce the transport model for CRs in SBNi which essentially provides the CR spectra used to study non-thermal emissions and also evaluate the ionization rates. The relevant radiative and ionization processes are introduced in Section \ref{sec:nonthermal}. We then perform a fit of the transport model to non-thermal emission data from the nuclei of NGC 253, M82, and Arp 220 to derive the CR spectra in these systems which can be employed to predict ionization rates $\zeta(\rm H_{2})$ for molecular clouds of different column densities. Our predicted values of $\zeta(\rm H_{2})$ for NGC 253 seem to be, in most cases, a few times smaller than the values recently inferred by \citetalias{holdship2022} and \citetalias{behrens2022} using molecular line observations (see Section \ref{sec:SBNi}). We also discuss in Section \ref{sec:discussion} the potential implications of this discrepancy. 

\section{Cosmic-ray transport in starburst nuclei}
\label{sec:transport}

We will follow the approach presented in \citet{peretti2019} and adopt the following transport equation to describe the differential number density $f_i(E)$ for CRs of species $i={\rm p}$ (protons) or $i={\rm e}$ (electrons) with kinetic energy $E$ in SBNi  
\begin{eqnarray}
\frac{f_i(E)}{\tau_{\rm adv}(E)}+\frac{f_i(E)}{\tau_{{\rm diff},i}(E)}-\frac{\partial}{\partial E}\left[b_i(E)f_i(E)\right]=Q_i(E),
\label{eq:transport}
\end{eqnarray}
where $Q_i(E)$ is the injection spectrum of CRs from SNRs (or as secondary and tertiary products from CR interactions with the ISM of SBNi), $b_i(E)$ is the energy loss rate due to interactions of CRs with SBNi's materials, $\tau_{\rm adv}$ and $\tau_{{\rm diff},i}(E)$ are respectively timescales for the escape of CRs from the SBNi due to advection and diffusion. 

The advection timescale is simply $\tau_{\rm adv}=R/u_w$ where $R$ is the radius of the SBNi and $u_w$ is the speed of galactic winds in SBNi. The diffusion timescale could be estimated as $\tau_{{\rm diff},i}(E)=R^2/D_i(E)$ where $D_i$ is the diffusion coefficient of CRs in SBNi. Here, we adopt Model A from \citet{peretti2019} which assumes the diffusive motion of particles to be induced by magnetic turbulence following a Kolmogorov power spectrum and the diffusion coefficient scales as     
\begin{eqnarray}
D_i(E)=\frac{L_0v_i}{2\eta_B}\left(\frac{r_{L,i}(E)}{L_0}\right)^{\frac{1}{3}},\label{eq:diffusion_coefficient}
\end{eqnarray}
where $L_0=1$ pc is the injection scale of turbulence, $\eta_B=\delta B^2/B^2=1$ is the ratio between the variance of turbulent magnetic fields $\delta B^2$ and the ordered field strength squared $B^2$, $v_i$ is the speed of CRs of species $i$ with kinetic energy $E$, and $r_{L,i}(E)$ is the Larmor radius of CRs of species $i$ with kinetic energy $E$. A complete list of energy loss processes for both protons and electrons can be found in \citet{schlickeiser2002} (see also \citealt{evoli2017}, and \citealt{peretti2019}). We note that the energy loss rates depend on many parameters characterizing the ISM of SBNi including magnetic field strength $B$, ISM density $n_{\rm ISM}$, electron density $n_{\rm e}$, electron temperature $T_{\rm e}$, and interstellar radiation field (see Eq. \ref{eq:photons} below). 

Concerning the injection spectrum, we consider CR protons injected only from SNRs and their injection spectrum could be modelled as a power law in momentum $p$  
\begin{eqnarray}
Q_{\rm p}(E)&=&Q_{\rm SNR,p}(E)\n\\
&=&\frac{\mathcal{R}_{\rm SNR}\xi_{\rm CR,p}E_{\rm SNR}}{V\Lambda \mpr^2 c^3 v_{\rm p}}\left(\frac{p}{\mpr c}\right)^{2-\alpha}\exp\left(-\frac{p}{p_{\rm max,p}}\right),
\label{eq:QSNRp}
\end{eqnarray}
where $\mathcal{R}_{\rm SNR}$ is the rate of supernova explosions within the SBNi, $\xi_{\rm CR,p}$ is the fraction of supernova explosion kinetic energy converted into CR kinetic energy (also referred to as the acceleration efficiency), $E_{\rm SNR}\simeq 10^{51}$ erg is the typical kinetic energy of supernova explosions, $p_{\rm max,p}\simeq 10^{17}$ eV/c is the cut-off momentum for CR protons accelerated from SNRs, $\mpr$ is proton mass, $c$ is the speed of light, $V=4\pi R^3/3$ is the volume of the SBNi, and $\Lambda=\int_{x_{\rm min}}^{\infty} x^{2-\alpha}\exp\left(-\frac{x}{x_{\rm max}}\right)\left(\sqrt{x^2+1}-1\right)\df x$. Note that the normalization factor $\Lambda$ ensures that a fraction $\xi_{\rm CR,p}$ of supernova explosion kinetic energy is converted into CR kinetic energy, meaning 
\begin{eqnarray}
\int^\infty_0EQ_{\rm SNR,p}(E)\df E = \frac{\mathcal{R}_{\rm SNR}\xi_{\rm CR,p}E_{\rm SNR}}{V}.
\end{eqnarray}
Throughout this work, we will assume $\xi_{\rm CR,p}=0.1$ which is, roughly speaking, around the typical value adopted for Galactic SNRs in models attempting to fit CR spectra in the local ISM \citep[see e.g.][]{evoli2019,phan2021,cristofari2021,merstch2021}.

For CR electrons, three different types of injection are taken into account
\begin{eqnarray}
Q_{\rm e}(E)=Q_{\rm SNR,e}(E)+Q_{\rm sec}(E)+Q_{\rm ter}(E),
\end{eqnarray}
where $Q_{\rm SNR,e}(E)$ is the injection spectrum of CR electrons from SNRs, $Q_{\rm sec}(E)$ is the injection spectrum of secondary electrons (and positrons) from the decay of $\pi^\pm$ produced in proton-proton interactions, and $Q_{\rm ter}(E)$ is the injection spectrum of tertiary electrons (and positrons) created in interactions between CR-induced gamma-rays and low-energy photons in the ISM of SBNi. In the following, we assume also a power law injection spectrum for CR electrons from SNRs $Q_{\rm SNR,e}(E) \propto \xi_{\rm CR,e}p^{2-\alpha}\exp(-p/p_{\rm max,e})$ with $\xi_{\rm CR,e}=0.01$ and $p_{\rm max,e}\simeq 10^{13}$ eV/c. For secondary electrons, we adopt the approach presented in \citet{kelner2006} to model the injection spectrum of secondary electrons as follows
\begin{eqnarray}
Q_{\rm sec}(E)=2\int^\infty_{E}\frac{cn_{\rm ISM}}{K_\pi}\sigma_{\rm pp}\left(\frac{E_\pi}{K_\pi}\right)f_{\rm p}\left(\frac{E_\pi}{K_\pi}\right)\tilde{f}_e\left(\frac{E}{E_\pi}\right)\frac{\df E_\pi}{E_\pi},
\end{eqnarray}
where $K_\pi\simeq 0.17$ is the fraction of kinetic energy transferred from the parent proton to the single pion, $\sigma_{\rm pp}(E)$ is the total inelastic cross-section for interactions between CR protons with kinetic energy $E$ and protons in the ISM of SBNi \citep{kafexhiu2014}, and $\tilde{f}_{\rm e}(E/E_\pi)$ is defined as in \citet{kelner2006} (see also Appendix B of \citealt{peretti2019}). Concerning tertiary electrons, the injection spectrum is
\begin{eqnarray}
Q_{\rm ter}(E)=\frac{c}{R}f_{\rm nth}(E_\gamma=E)\left[1-{\rm exp}\left(-\tau_{\gamma\gamma}(E_\gamma=E)\right)\right],
\end{eqnarray}
where $f_{\rm nth}(E)$ is the differential number density of non-thermal photons induced by CRs and $\tau_{\gamma\gamma}(E)$ is the opacity of gamma-rays due to interactions with low-energy photons in the ISM of SBNi which could be estimated as follows
\begin{eqnarray}
f_{\rm nth}(E_\gamma)&\simeq&\frac{R}{c}\left\{\int^\infty_0\df E f_{\rm p}(E) \epsilon_{\rm PPI}(E,E_\gamma)\right.\n\\
&&\left.+\int^\infty_0\df E f_{\rm e}(E) \left[\epsilon_{\rm BRE}(E,E_\gamma)+\epsilon_{\rm ICS}(E,E_\gamma)\right.\right.\n\\
&&\left.\qquad\qquad\qquad\qquad\qquad\qquad+\left.\epsilon_{\rm SYN}(E,E_\gamma)\vphantom{^{a}}\right]\vphantom{\int^\infty_0}\right\}\n\\
\label{eq:f_gamma}\\
\tau_{\gamma\gamma}(E_\gamma)&=&\int \df E'_\gamma f_{\rm ISRF}(E'_\gamma)\sigma_{\gamma\gamma}(E_\gamma,E'_\gamma)R.\label{eq:tau_gg}
\end{eqnarray}
Here, we have introduced the volume emissivities for four main processes for non-thermal emissions induced by CRs $\epsilon_{\rm PPI}$, $\epsilon_{\rm BRE}$, $\epsilon_{\rm ICS}$, and $\epsilon_{\rm SYN}$ which are respectively proton-proton interactions, bremsstrahlung radiation, inverse Compton scattering, and synchrotron radiation. These processes will be discussed in more detail in the next section. In principle, the source spectrum of tertiary electrons should depend on bremsstrahlung radiation and inverse Compton scattering induced by CR electrons themselves which makes the problem non-linear. We shall see later that the observed gamma-ray spectrum of SBNi might be dominated by the hadronic gamma-ray component and, thus, one can neglect the non-linearity in the CR transport equation for electrons by considering tertiary electrons coming from hadronic gamma-rays only. As for the opacity of gamma-rays in SBNi, we have introduced also the gamma-gamma interaction cross-section $\sigma_{\gamma\gamma}(E,E_{\rm ph})$ (\citealt{aharonian2013}) and the differential number density of interstellar photons $f_{\rm ISRF}(E_{\rm ph})$ which shall be introduced in more details in the next section (see Eq. \ref{eq:photons}).   

Having discussed all the relevant ingredients, we could essentially proceed to the solution of the transport equation applicable for both CR protons and electrons
\begin{eqnarray}
f_i(E)&=&\frac{1}{b_i(E)}\int^{E_{\rm max}}_E\df E'Q_i(E')\n\\
&&\times\,{\rm exp}\left[-\int_{E}^{E'} \frac{\df E''}{b_i(E'')}\left(\frac{1}{\tau_{\rm adv}}+\frac{1}{\tau_{\rm diff,i}(E'')}\right)\right].\label{eq:solution-transport}
\end{eqnarray}
Note that the CR differential number density could be related to the CR spectra or CR flux as follows $j_i(E)=f_i(E)v_i/(4\pi)$. 

\section{Non-thermal emissions and ionization rates in starburst nuclei}
\label{sec:nonthermal}

\subsection{Cosmic-ray induced gamma-rays and X-rays}
\label{sec:radiative}
As mentioned in the previous section, we will focus mostly on CR induced gamma-rays from the decay of $\pi_0$ in proton-proton interactions, bremsstrahlung radiation, and inverse Compton scattering. In the following, we shall briefly discuss the volume emissivities for these processes (interested readers could find some more details in \citealt{peretti2019}).  

Concerning $\pi_0$ decay, we shall follow the approach as presented in \citet{kelner2006} but adopt the differential cross-section for proton-proton interactions $\df\sigma_{\rm pp}/\df E_\gamma$ and the nuclear enhancement factor $\varepsilon_{\rm n}$ (to correct for gamma-rays induced by CR nuclei) from \citet{kafexhiu2014}. The volume emissivity could be modelled as follows
\begin{eqnarray}
\epsilon_{\rm PPI}(E,E_\gamma)=n_{\rm ISM} v_{\rm p} \varepsilon_{\rm n}(E)\frac{\df \sigma_{\rm pp}(E,E_\gamma)}{\df E_\gamma}.
\end{eqnarray}
Note that the differential cross-section will be non-zero only for CR protons with kinetic energy satisfying $E\leq E_\gamma+m_\pi^2c^4/(4E_\gamma)$.

As for bremsstrahlung radiation, the volume emissivity could be estimated following \citet{schlickeiser2002}
\begin{eqnarray}
\epsilon_{\rm BRE}(E,E_\gamma)=n_{\rm ISM}c\frac{\sigma_{\rm BRE}(E,E_\gamma)}{E_\gamma},
\end{eqnarray}
where the cross-section for bremsstrahlung radiation $\sigma_{\rm BRE}(E,E_\gamma)$ (see Chapter 4 of \citealt{schlickeiser2002} or \citealt{baring1999} for more details).

Another important process for gamma-rays induced by CR electrons is inverse Compton scattering where low-energy photons in the ISM are scattered by CR electrons and are boosted to higher energy. Modelling this process, thus, requires some knowledge of the low-energy interstellar photons especially in the far infrared to optical energy range. We shall follow \citet{peretti2019} and consider the interstellar radiation field made up mostly of four components: far-infrared (FIR), mid-infrared (MIR), near-infrared (NIR), and optical (OPT). The differential number density of each component could be modelled as follows 
\begin{eqnarray}
f_{\rm rad}(E_\gamma)=\frac{U_{\rm rad}(k_{\rm B}T_{\rm rad})^{-2}}{\Gamma(4+\sigma)\zeta(4+\sigma)\left({\rm e}^{{\frac{E_\gamma}{k_{\rm B}T_{\rm rad}}}}-1\right)}\left(\frac{E_\gamma}{k_{\rm B} T_{\rm rad}}\right)^{2+\sigma} \label{eq:thermal-emission}
\end{eqnarray}
where ${\rm rad=FIR}$, MIR, NIR, or OPT, $U_{\rm rad}$ is the energy density of the respective component, $T_{\rm rad}$ is the effective photon temperature of the respective component, $\Gamma(x)$ and $\zeta(x)$ are respective the Gamma and Riemann zeta functions, $\sigma$ is a spectral index which shall be set to be $\sigma=0$ for OPT and $\sigma\simeq 1.3$ for the rest. In fact, the differential number density would become that of black-body radiation for the case where $\sigma=0$. Note also that we have used the notation $E_\gamma$ to indicate the photon energy for both the thermal and non-thermal photons in all energy domains (from radio to gamma-ray). The differential number density of interstellar photons is then written as follows 
\begin{eqnarray}
f_{\rm ISRF}(E_\gamma)=\sum_{\rm rad}f_{\rm rad}(E_\gamma)
\label{eq:photons}
\end{eqnarray}
where the sum is performed over all the components of the interstellar radiation field as mentioned above. The volume emissivity of inverse Compton scattering is then~\citep{1970RvMP...42..237B}
\begin{eqnarray}
&&\epsilon_{\rm ICS}(E,E_{\rm\gamma})=\int^\infty_0\df E'_\gamma f_{\rm ISRF}(E'_\gamma)c\frac{\sigma_{\rm ICS}(E,E_\gamma)}{E'_\gamma},\n\\
&&\hphantom{\epsilon_{\rm ICS}(E,E_{\rm\gamma})}\simeq\sum_{\rm rad}\frac{U_{\rm rad}}{(k_B T_{\rm rad})^2}c\sigma_{\rm ICS}(E,E_\gamma).
\end{eqnarray}
where $\sigma_{\rm ICS}(E,E_\gamma)$ is the cross section for the inverse Compton scattering (see Chapter 4 of \citealt{schlickeiser2002} for more details). It is worth mentioning also that this process also has a threshold energy meaning that the emissivity is non-zero only for CR electrons with kinetic energy $E\gtrsim E_{\gamma}\left[1+\sqrt{1+m_{\rm e}^2c^4/(E_\gamma k_{\rm B}T_{\rm rad})}\right]$.  

In starburst galaxies, non-thermal radio and X-ray emissions can also come from synchrotron radiation induced by CR electrons. The volume emissivity of synchrotron radiation could be written as \citep{zirakashvili2007}
\begin{eqnarray}
\epsilon_{\rm SYN}(E,E_\gamma)=\frac{\sqrt{3}e^3(1+\eta_B)B^2}{m_ec^2 h E_\gamma}R\left(\frac{E_\gamma}{E_{\gamma,c}}\right),
\end{eqnarray}
where $B$ is the field strength of the magnetic field in the starburst nucleus, $h$ is the Planck constant, $R(x)=1.81e^{-x}/\sqrt{x^{-2/3}+1.33}$, and $E_{\gamma,c}\simeq\frac{3hE_e^2 }{2m_e^2c^4 v_e}\frac{e\sqrt{1+\eta_B}B}{2\pi m_e c}$. 

The flux of photons (for both thermal and non-thermal emissions) from these galaxies is then given as
\begin{eqnarray}
\phi(E_\gamma)&\simeq& \frac{1}{3}\left[f_{\rm nth}(E_\gamma)+f_{\rm ISRF}(E_\gamma)\right]c\left(\frac{R}{d_{\rm gal}}\right)^2\n\\
&&\times \frac{1-\exp\left[-\tau_{\rm ff}(E_\gamma)-\tau_{\gamma\gamma}(E_\gamma)\right]}{\tau_{\rm ff}(E_\gamma)+\tau_{\gamma\gamma}(E_\gamma)}\exp\left[-\tau_{\rm EBL}(E_\gamma)\right]\n\\
\end{eqnarray}
where $f_{\rm nth}(E_\gamma)$ and $f_{\rm ISRF}(E_\gamma)$ are respectively defined in Eq. \ref{eq:f_gamma} and Eq. \ref{eq:photons}, $d_{\rm gal}$ is the distance between the galaxy of interest and the Milky Way, $\tau_{\rm ff}(E_\gamma)$ is the opacity due to free-free absorption relevant in the radio domain (see Appendix \ref{appendixA}), and $\tau_{\gamma\gamma}(E_\gamma)$ and $\tau_{\rm EBL}(z_{\rm SBN},E_\gamma)$ are the opacities due to interactions of high-energy photons respectively with ISRF of the SBN and with the extragalactic background light relevant in the gamma-ray domain. 
We have defined $\tau_{\gamma\gamma}(E_\gamma)$ in Eq. \ref{eq:tau_gg} and adopt the analytic form of $\tau_{\rm EBL}(E_\gamma)$ as presented in Appendix C of \citet{peretti2019}. Note that the factor $1/3$ is due to the effective spherical shape assumed for the SBNi. The thernal and non-thermal flux shall be fitted later with observational data to obtain the source and transport parameters of CRs in SBNi. 

\begin{figure}
    \centering
	\includegraphics[width=3.5in]{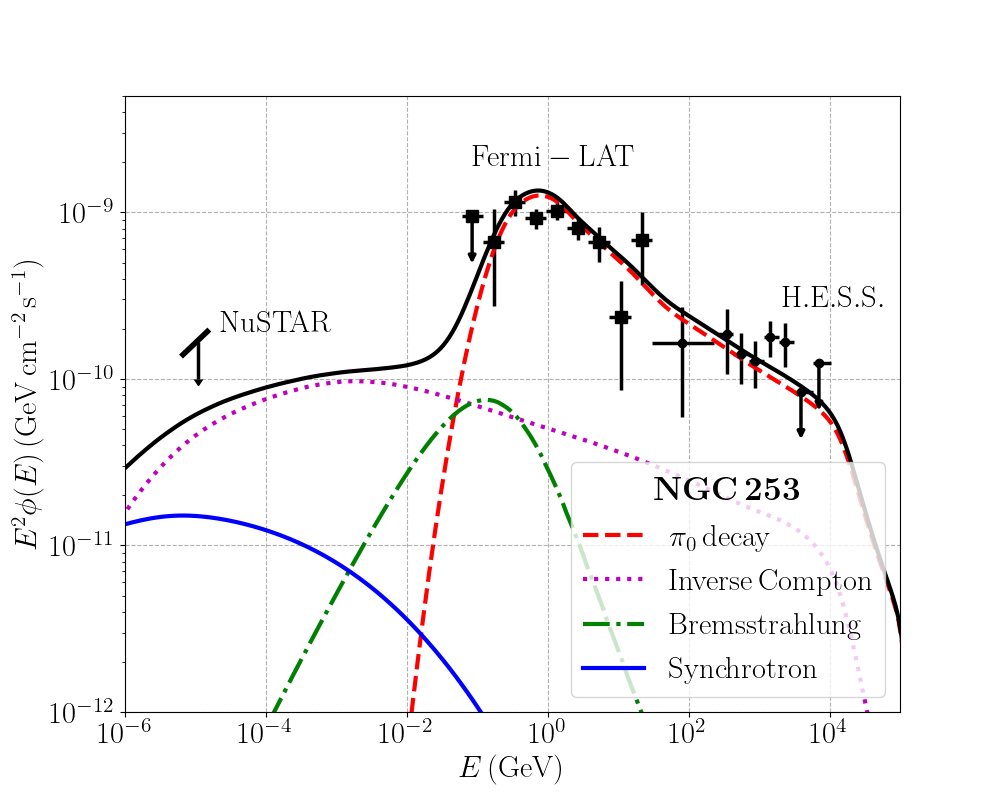}
    \caption{Non-thermal emissions from the SBN of NGC 253 from hard X-ray to TeV gamma-ray domains from $\pi_0$ decay (dashed red line), inverse Compton scattering (dotted magenta line), bremsstrahlung radiation (dash-dotted green line), and synchrotron radiation (solid blue line). The flux is fitted to gamma-ray data from Fermi-LAT and H.E.S.S. \citep{abdalla2018} and upper limits from NuSTAR \citep{wik2014}.}
    \label{fig:gamma_spectrum}
\end{figure}

\subsection{Cosmic-ray induced ionization rates}
\label{sec:ionization}

Once the parameters determining CR spectra in SBNi have been fitted using gamma-ray and X-ray data, we could proceed to predict the ionization rate in molecular clouds that are embedded within these systems. We note, however, that the CR-induced ionization rate should actually vary for clouds with different column densities. Self-consistent predictions of this quantity require a model to describe the transport of CRs into molecular clouds. In fact, CR transport might be {\it ballistic} (gyrating along magnetic field lines, \citealt{padovani2009}) or {\it diffusive} (executing random walks along magnetic field lines, \citealt{morlino2015,ivlev2018,phan2018,owen2021}) depending on the geometry of magnetic fields threading the clouds. In particular, the appropriate model for the transport of CRs into a cloud can be chosen by comparing the cloud's size $L_{cl}$ to the magnetic field coherence length in the ISM $l_c$ (see \citealt{phan2023} for more an extended discussion on different models). The diffusive or ballistic model should be preferred for $L_{cl}\ll l_c$ or $L_{cl}\simeq L_c$ respectively.

In SBNi, the value of $l_c$ is, in fact, not very well known. However, if we assume that the large scale magnetic turbulence in these systems are also generated by supernova explosions similar to that in the Milky Way \citep{berezinskii1990,blasi2013,evoli2018}, then the value of $l_c$ should be comparable to the typical sizes of SNRs. We could take as a rough estimate for $l_c$ the size of an SNR at the end of the Sedov-Taylor phase which, for a typical density of around $200$ cm$^{-3}$ in SBNi, is about 6 pc (assuming a typical ejecta mass of around $1\,M_\odot$). Such a coherence length should be comparable to size of clouds in the SBNi of interest.
It is for this reason that we shall adopt the {\rm ballistic} model where the transport of CRs inside clouds is one-dimensional and the average CR spectra in a cloud of size $L_{cl}$ could be estimated as follows
\begin{eqnarray}
    &&f_{i,cl}(E)=\int^{L_{cl}}_0\frac{\df x}{2 L_{cl}}\n\\
    &&\hphantom{f_{i,cl}(E)}\times\left[\frac{f_i(E_{01})b_{i,cl}(E_{01})}{b_{i,cl}(E)}+\frac{f_i(E_{02})b_{i,cl}(E_{02})}{b_{i,cl}(E)}\right]\n\\\label{eq:f_a}
\end{eqnarray}
where $f_i(E)$ is the differential number density for CRs of species $i$ ($i={\rm p}$ or $i={\rm e}$ respectively for protons and electrons) in the ISM of SBNi as obtained from Eq. \ref{eq:transport}, $b_{i,cl}(E)\simeq n_{cl} b_{i}(E)/n_{\rm ISM}$ is the energy loss rate of CRs inside clouds ($n_{cl}$ is the gas density inside the cloud), $E_{01}=E_0(x,E)$ and $E_{02}=E_0(L_{cl}-x,E)$ where the function $E_0$ (depending on the species of CRs considered) is the initial energy of CRs as they enter the cloud and is obtained by solving the following equation 
\begin{eqnarray}
    x=\int^{E_0}_{E}\frac{v_i\df E}{b_i(E)}.
\end{eqnarray}
In fact, it can be shown that the average differential number density as defined in Eq. \ref{eq:f_a} depends only on the total column density of the cloud which, for dense molecular clouds, is $n_{cl}L_{cl}\simeq 2N({\rm H}_2)$. Thus, given the CR flux in the ISM of SBNi, we could predict the ionization rate inside clouds as a function of the H$_2$ column density $N({\rm H}_2)$.

The H$_2$ ionization rate induced by CR protons and electrons could be obtained as in \cite{padovani2009} \citep[see also][]{chabot2016,phan2018,recchia2019}:
\begin{eqnarray}
\zeta_i(\text{H}_2)&=&\int^{\infty}_{I({\rm H}_2)} f_{i,cl}(E) \, v_i \left[1+\phi_i(E)\vphantom{^{\frac{•}{•}}}\right]\sigma^i_{\text{ion}}(E)\df E\label{eq:zeta_i}
\end{eqnarray}
where $\sigma^{i}_{\text{ion}}$ is the ionization cross-section of CR species $i$, $\phi_i(E)$ are the average secondary ionization per primary ionization computed as in \cite{krause2015} (see also \citealt{padovani2009,ivlev2021}), and $I({\rm H}_2)=15.603$ eV is the ionization potential of H$_2$. It should be noticed that, following \cite{krause2015}, the two ionization cross-sections are considered in the computation with fully relativistic corrections. The total CR-induced ionization rate is then 
\begin{eqnarray}
\zeta({\rm H}_2)\simeq 1.5\zeta_{\rm p}({\rm H}_2)+\zeta_{\rm e}({\rm H}_2).\label{eq:zeta}
\end{eqnarray}

It is worth mentioning also that we will ignore the ionization rate induced by X-rays. In fact, X-rays with energy from $\sim 1$ to 10 keV from synchrotron radiation and inverse Compton scattering of CR electrons in nuclei of starburst galaxies could also contribute to the ionization rate but we have checked that $\zeta_{\rm X-ray}({\rm H}_2)\lesssim 10^{-17}$ s$^{-1}$ for clouds with $N({\rm H}_2)\gtrsim 10^{23}$ cm$^{-2}$. 

Having discussed all the relevant radiative processes in the previous subsection, we could now apply them to derive the CR spectra in several prototypical nearby SBNi, namely NGC 253, M82, and Arp 220 by fitting non-thermal emission data from these objects. This should then allow us to predict ionization rates within these SBNi using the ballistic model.

\begin{figure}
	\includegraphics[width=3.5in]{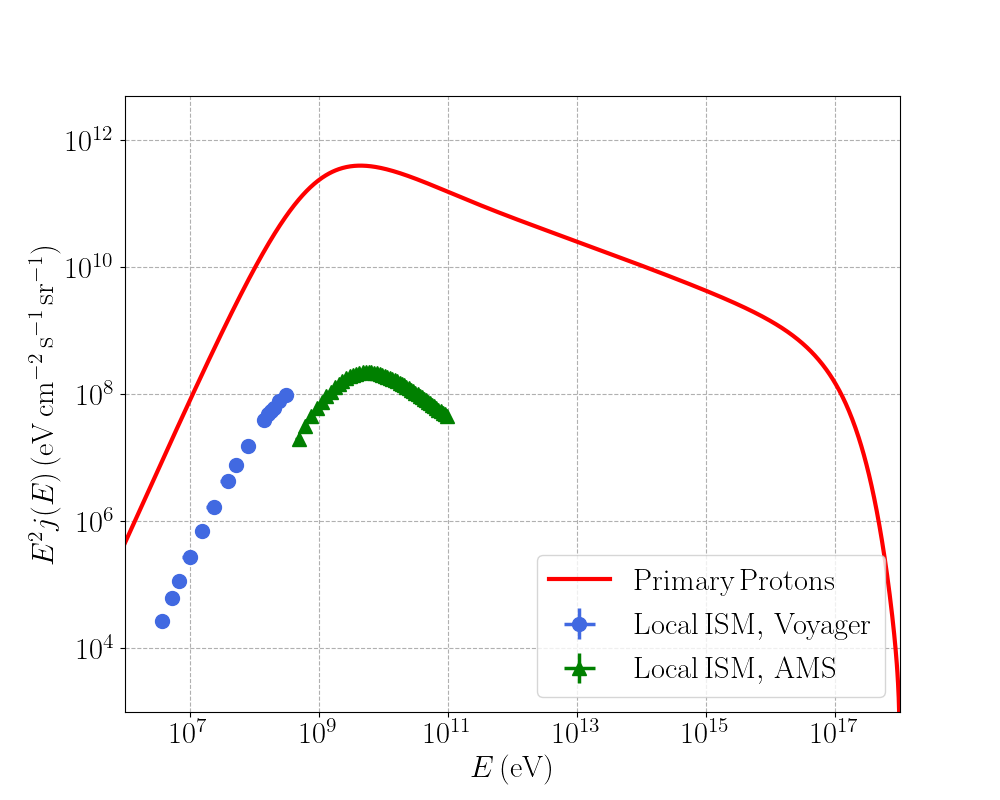}\\
 	\includegraphics[width=3.5in]{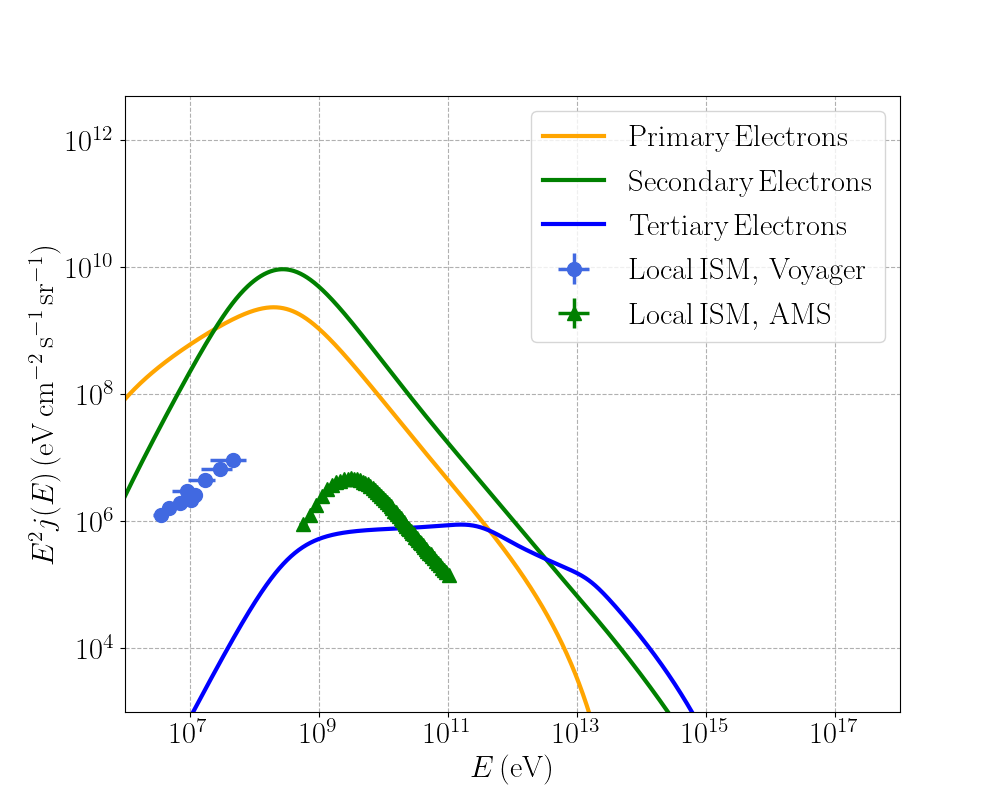}
    \caption{Spectra of CR protons (upper panel) and electrons (lower panel) in the SBN of NGC 253 obtained using the source and transport parameters adopted for the fit of non-thermal emissions. Data of CR spectra observed in the local ISM is also presented for comparison. For CR electrons, we present different components including primary electrons from SNRs (orange line), secondary electrons from decays of $\pi^\pm$ (green line), and tertiary electrons from interactions between CR-induced gamma-rays and interstellar radiation (blue line).}
    \label{fig:CR_spectra}
\end{figure}

\section{Implications for known starburst nuclei}
\label{sec:SBNi}

\subsection{Starburst nucleus of NGC 253}

NGC 253 is a spiral galaxy located at a distance of $d_{\rm gal}\simeq 3.8$ Mpc \citep{rekola2005,dalcanton2009} making it one of the closest objects to be classified as a starburst galaxy in the southern sky. This starburst galaxy is believed to have a star formation rate (SFR) of about $5\,M_\odot\,{\rm yr}^{-1}$ which is a few times higher than that of the Milky Way. About $70\%$ of the star-forming activity occurs, however, in the starburst nucleus region \citep{melo2002}. As a consequence, the SBN of NGC 253, which is of size $R\simeq 100 \, \text{pc}$, has a relatively high supernova rate with $\mathcal{R}_{\rm SNR}\simeq 0.03$ yr$^{-1}$ comparable to that of the entire Milky Way \citep{engelbracht1998}.      

\begin{figure}
    \centering
	\includegraphics[width=3.5in]{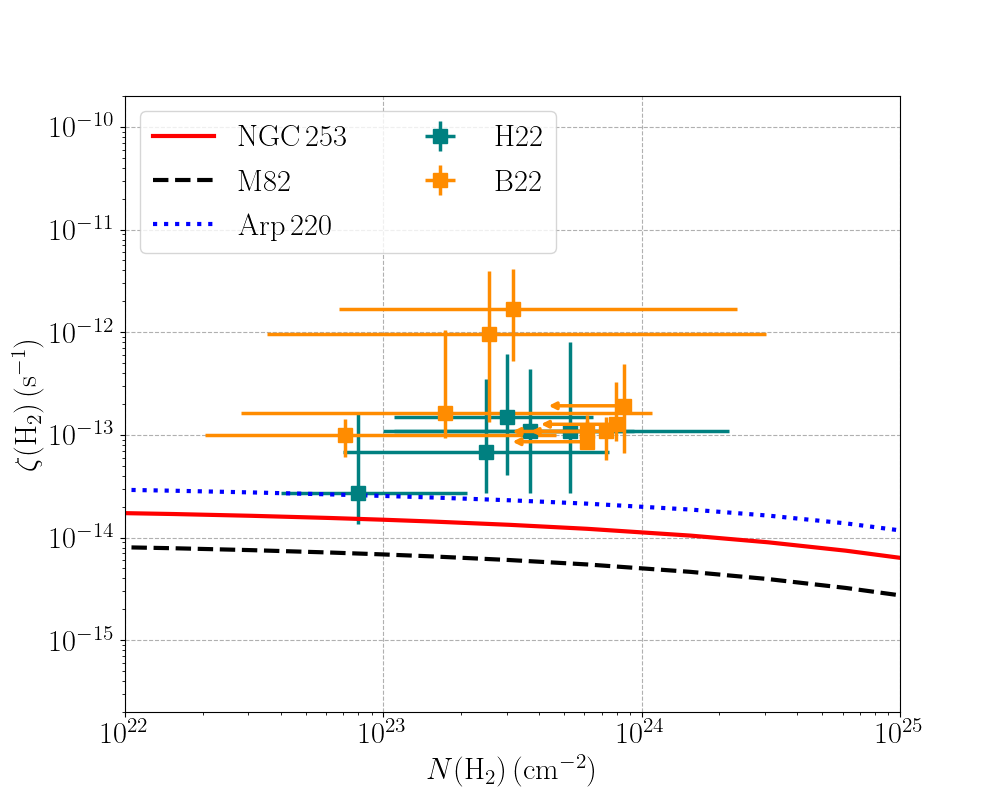}
	\includegraphics[width=3.5in]{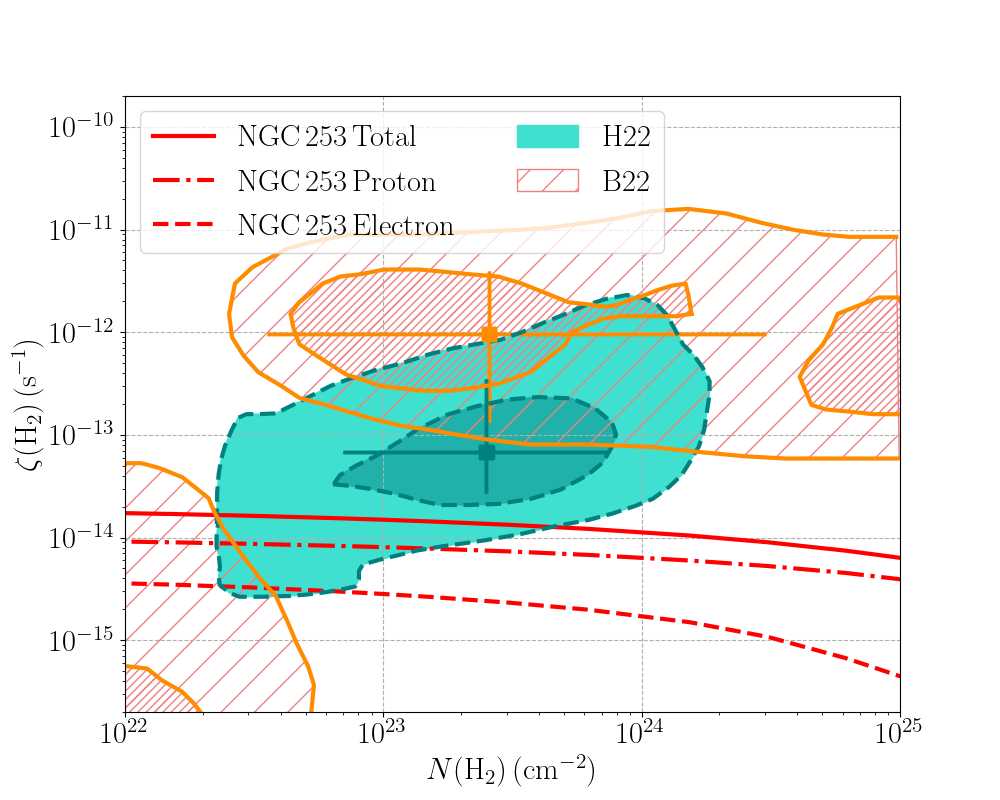}
    \caption{{\it Upper}: Cosmic-ray induced ionization rate expected in the SBN of NGC 253 (solid red line), M82 (dotted blue line), and Arp220 (dashed black line). Inferred values of the ionization rates for a few molecular clouds in the central region of NGC 253 from \citetalias{holdship2022} (green squares) and \citetalias{behrens2022} (orange squares) are also overlaid for comparison. {\it Lower}: Cosmic-ray induced ionization rate expected for NGC 253 is shown again together with the $1\sigma$ and $2\sigma$ contours of the ionization rate inferred for a particular molecular cloud in the SBN, referred to as GMC 6 in the analyses of both \citetalias{holdship2022} and \citetalias{behrens2022}. Seperate contributions of CR protons (dotted red line) and electrons (dashed red line) to the total ionization rate in the SBN of NGC 253 are also presented.} 
    \label{fig:ion}
\end{figure}

Interestingly, the nucleus of NGC 253 has been observed in gamma-rays both in the GeV and TeV energy ranges with Fermi-LAT and H.E.S.S. telescopes (\citealt{abdo2010,abdalla2018}, see also \citealt{abramowski2012}). There exist also upper limits in the energy range around 10 keV by the X-ray telescope NuSTAR \citep{wik2014}. Recently, several molecular clouds in the central region of this SBN have also been targeted to study ionization rates by the ALCHEMI Collaboration using chemical surveys performed with ALMA data \citepalias{holdship2022,behrens2022}. We shall now estimate the ionization rates in the SBN of NGC 253 using non-thermal emissions and compare them to the ones derived from molecular line observations.  

All the parameters relevant for modelling both thermal and non-thermal emissions from the SBN are presented in Table \ref{tab:parameters}. We have fixed the distance, the SBN size, and the galactic wind speed as in \citet{peretti2019} which are also motivated from several independent observations. The remaining parameters are fitted to data and upper limits from radio to gamma-ray and the results are shown in Fig. \ref{fig:gamma_spectrum}.
The parameter choice leads to the expected flux being dominated by gamma-rays from $\pi_0$ decay above about 100 MeV. Gamma-rays and X-rays from bremsstrahlung, inverse Compton scattering, and synchrotron radiation induced by CR electrons, on the other hand, contribute significantly to the flux below about 100 MeV with the expected flux consistent with the upper limit set by NuSTAR. Interestingly, the choice of the galactic wind speed and the diffusion coefficient as in Eq. \ref{eq:diffusion_coefficient} mean that the transport of CRs in this system is likely dominated by energy loss (calorimetric system) given a typical ISM density $n_{\rm ISM}\gtrsim 100$ cm$^{-3}$. In this case, the solution of the transport equation (see Eq. \ref{eq:solution-transport}) simplifies to $j(E)\sim Q(E)E/b(E)\sim \xi_{\rm CR,p}\mathcal{R}_{\rm SNR}E^{2-\alpha}/(n_{\rm ISM}R^3)$ for $E\gtrsim 300$ GeV. If we notice also that the hadronic gamma-ray flux at energy $E_\gamma$ is also, roughly speaking proportional to the CR proton spectrum at $E\simeq 10 E_\gamma$, the SBN gas mass and the inverse of distance squared, we can show that $\xi_{\rm CR,p}\mathcal{R}_{\rm SNR}E^{2-\alpha} d^2 \sim \phi(E_\gamma=E/10)$ (see e.g. \citealt{abramowski2012} for similar discussions). Such a relation means that, given fixed values of $\xi_{\rm CR,p}$ and $d$, the index of the CR injection spectrum $\alpha$ and the supernova rate $\mathcal{R}_{\rm SNR}$ can be constrained respectively by the spectral index and the normalization of the high-energy gamma-ray spectrum. More importantly, the exact shape of the predicted gamma-ray spectrum around GeV energy is determined by the values of the ISM density $n_{\rm ISM}$. For NGC 253, we have found that $\alpha=4.3$, $\mathcal{R}_{\rm SNR}=0.03$ yr$^{-1}$, and $n_{\rm ISM}=170$ cm$^{-3}$. The value of $\mathcal{R}_{\rm SNR}=0.03$ yr$^{-1}$ is, in fact, consistent with values derived from \citet{engelbracht1998} using spectroscopic data. We note, however, that uncertainties on gamma-ray data mean that this value can be uncertain within a factor of two. Also, the gas mass of the SBN derived from $n_{\rm ISM}$ and $R$ is about $7\times 10^{7} M_\odot$ which is within the uncertainty range indicated by other estimates using molecular line observations \citep{bradford2003}.  

We present also the fit results in the frequency range from radio to optical in Fig. \ref{fig:radio} of Appendix \ref{appendixA} with data from various observations retrieved from the {\it NASA/IPAC Extragalactic Database}\footnote{https://ned.ipac.caltech.edu/}. 
It is clear that synchrotron radiation becomes relevant in the frequency range below a few tens GHz. Here, the magnetic field has been fitted to $B=120\,\mu{\rm G}$ which is comparable to the magnetic field obtained by assuming equipartition of energy density between CRs and magnetic field and, thus, quite conservative as a lower limit for the values of $B$. The fitted magnetic field strength also ensure that the synchrotron radiation from secondary CR electrons (fixed by the CR protons and the ISM density) do not surpass upper limits derived by \citet{williams2010} using data from the Allen Telescope Array around 1 GHz. In addition, we have chosen the electron acceleration efficiency $\xi_{\rm CR,e}=0.01$ as commonly adopted in studying Galactic CRs and this lead to a subdominant contribution of primary CR electrons in the radio domain compatible with the available upper limits. The tight constraints in the GHz domain, in fact, leave little room for increasing the value of $\xi_{\rm CR,e}$ much above the Galactic value.

The corresponding CR spectra derived from these fit parameters are also shown in Fig. \ref{fig:CR_spectra} with CR data from the local ISM overlaid for comparison. It is clear the CR spectra in the SBN of NGC 253 are many orders of magnitude larger than that in the local ISM, especially in the MeV to GeV energy range. Thus, we expect the CR-induced ionization rate to be also much larger than typical Galactic values.  

Using parameters obtained in the fit of non-thermal emissions, we could now predict the CR-induced ionization rate in the SBN of NGC 253. In this case, the ionization rate could reach about \mbox{$\zeta({\rm H}_2)\simeq 1.5\times 10^{-14} \, \text{s}^{-1}$} for clouds with column density \mbox{$N({\rm H}_2)=10^{23} \, \text{cm}^{-3}$}. The ionization rate for clouds of different column densities is shown in Fig. \ref{fig:ion} together with the inferred values of ionization rates in several giant molecular clouds from \citetalias{holdship2022} and \citetalias{behrens2022}. Separate contributions of CR protons and electrons to the total ionization rate in the SBN of NGC 253 are also shown in the lower panel of Fig. \ref{fig:ion}. Interestingly, the contributions from CR protons are always dominant over that of electrons. This result seems rather conservative given the tight constraints of CR electrons in the radio domain. It is clear the predicted ionization rate is lower than the values inferred from observations of molecular lines. The differences in most cases are, however, only a factor of a few to roughly one order of magnitude below the data points, except for two extreme cases from \citetalias{behrens2022} where the ionization rates reach a value of a few $10^{-12}$ s$^{-1}$. There might be many potential explanations for such a discrepancy which will be elaborated in Section \ref{sec:discussion}. At this point, we would like however to provide a short discussion on ionization rate data derived using chemical modeling for molecular line observations in comparison to our predicted ionization rate.  

Indeed, the cosmic ray ionization rates obtained by these studies are actually derived by fitting, or more precisely performing Bayesian inference for, a chemical model with a small number of parameters, including also $\zeta({\rm H_{2}})$ and $N({\rm H_{2}})$, to molecular line observations. This gives, in the end, a posterior distribution of all the parameters which, ultimately, allow us to quantity the values of $\zeta({\rm H_{2}})$ and $N({\rm H_{2}})$ together with their uncertainties. This procedure are adopted for both \citetalias{holdship2022} and \citetalias{behrens2022} and, in fact, the two analyses are performed for the same set of molecular clouds but study emissions from different molecules: \citetalias{holdship2022} focus on H$_3$O$^+$ and SO and \citetalias{behrens2022} examine HCN and HNC. In other words, each green data point in the upper panel of Fig. \ref{fig:ion} has a corresponding yellow data point and yet their values are different by a factor of a few to roughly one order of magnitude in most cases. This could mean that the quoted error bars from these data points might not fully reflect the uncertainties in chemical modelling adopted to derive the ionization rates. This is likely due to uncertainties intrinsic to: the observations (performed over clouds of almost 30 pc typical size, and sometimes possibly biased towards warm regions which is the case for H$_3$O$^+$, for which the lower-lying transitions are basically excluded from the analysis because satisfying fits cannnot be found), the radiative transfer of the species (with partial collisional processes implemented, neglecting collisions with electrons for instance), and decisions on chemical modelling (using a single point model for each cloud, excluding some reactions with unknown rates, approximating the sulfur depletion, and using equilibrium values) as well as on physical modelling (excluding the treatment of shocks or UV photons). In order to better illustrate these uncertainties, we provide in the lower panel of Fig. \ref{fig:ion} an example where our predictions of the ionization rates are overlaid with the 1$\sigma$ and 2$\sigma$ contours for a particular molecular cloud, referred to as GMC 6 by both \citetalias{holdship2022} and \citetalias{behrens2022} (see Table A1 of \citetalias{behrens2022} for the sky coordinate of this cloud). For the case of \citetalias{holdship2022}, our predicted ionization rates are actually within the 2$\sigma$ contour for this clouds if $N({\rm H}_2)\sim\times 10^{23}$ cm$^{-2}$. We have also checked that, roughly speaking, this is also true for all the data points from \citetalias{holdship2022}. Regarding the case of \citetalias{behrens2022}, the comparison is more complicated as there are, in fact, multiple regions of the plane $\zeta({\rm H}_2)$ - $N({\rm H}_2)$ where the chemical model gives good fit to molecular line data\footnote{The multimodal behaviour of the posterior distribution exists for several clouds studied by \citetalias{behrens2022} but not all of them.}. There are also regions of the posterior giving lower values for $\zeta({\rm H}_2)$ and $N({\rm H}_2)$ which are less likely but compatible with our predictions for clouds with low column densities.  

Nevertheless, we have illustrated that the ionization rate estimated from non-thermal emissions can be in most cases within a factor of a few different than that derived by molecular line observations. Thus, it can be used in complementarity to molecular line observations to provide more precise values for ionization rates which might be useful for chemical modeling of complex star-forming regions.      


\subsection{Starburst nuclei of M82 and Arp 220}
We can also apply the framework presented above to study the ionization rate in SBNi of M82 and Arp 220. These two starburst galaxies are also relatively nearby with high SFRs (and correspondingly high supernova rates) such that they are also visible by gamma-ray telescopes in the GeV and TeV energy ranges and by the X-ray telescope Chandra in the keV energy range. We will again model the underlying CR spectra which could account for the non-thermal emissions and employ these spectra to predict ionization rates in the SBNi. 

Concerning M82, this is a nearby starburst galaxy (about $3.9$ Mpc from the Milky Way, \citealt{sakai1999}) and it is very well known for hosting a galactic superwind with a wind speed reaching several hundred kilometers per second \citep{strickland2009}. This galaxy also has an active compact starburst nucleus (of size $R\simeq 150$ pc, \citealt{volk1996}) which is believed to form due to its interactions with the nearby spiral galaxy M81 \citep{yun1994}. The SBN has been inferred to have an SFR which is about 10 times higher than that of the Milky Way and the corresponding supernova rate in this compact starburst region is $\mathcal{R}_{\rm SNR}\simeq 0.1$-$0.3$ yr$^{-1}$ \citep{kronberg1985,fenech2008}. 

Since the distance, size, and supernova rate of this SBN are very similar to that of NGC 253, the nucleus of M82 is also expected to be a bright gamma-ray and X-ray source. In fact, it has been observed also in the GeV energy range by Fermi-LAT \citep{acero2015} and in the TeV energy range by VERITAS \citep{acciari2009}. Observations with several telescopes including Chandra, XMM-Newton, and NuSTAR also reveal a very high flux of hard X-ray around 10 keV from the M82 SBN \citep{strickland2007,ranalli2008,bachetti2014}. However, it has been argued by \citet{peretti2019} that there might exist unresolved X-ray sources in the SBN and, thus, the observed X-ray flux should be treated as an upper limit for the expected flux of non-thermal emissions induced by CRs.  

\begin{figure}
	\includegraphics[width=3.5in]{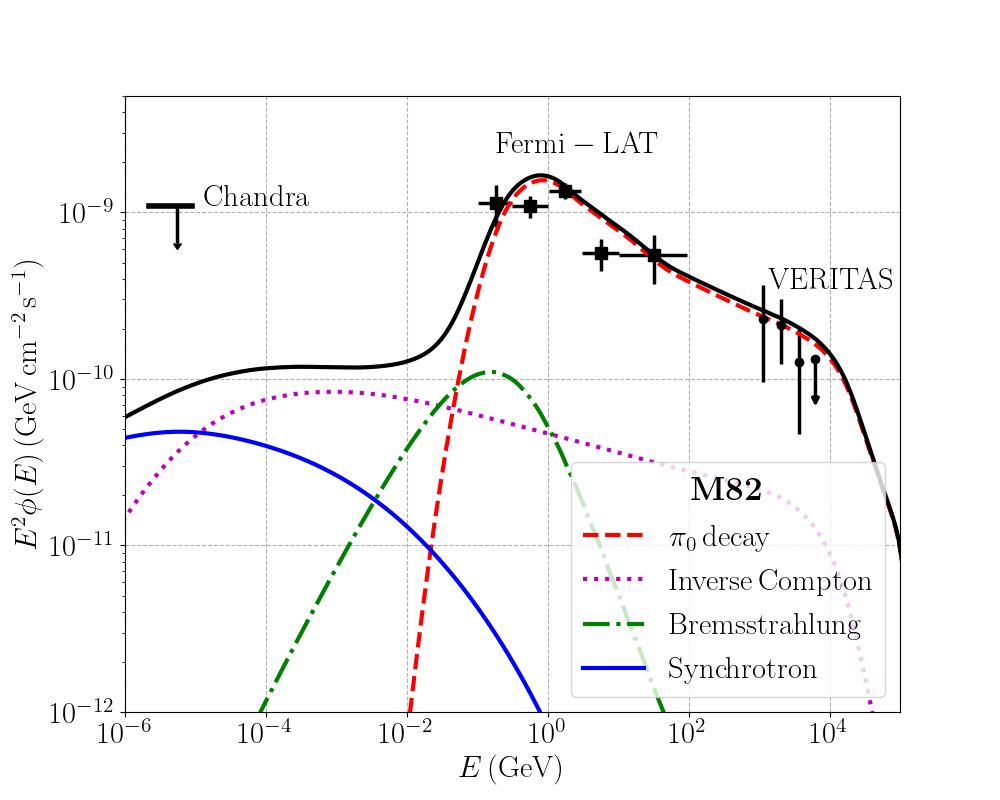}\\
	\includegraphics[width=3.5in]{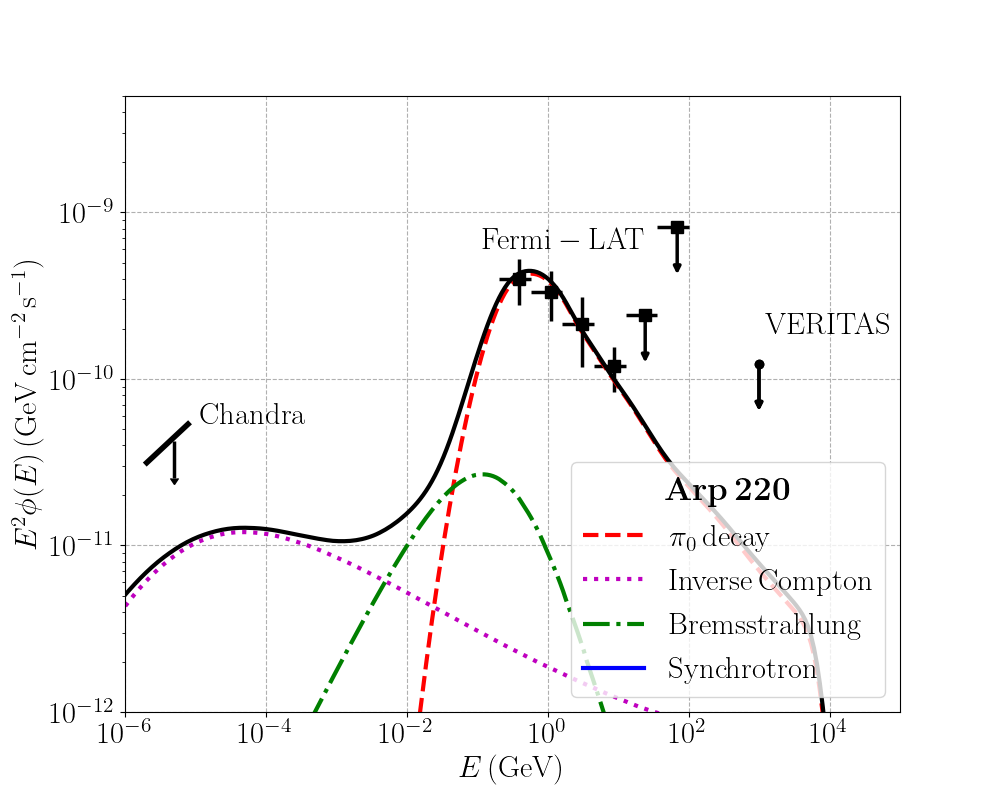}
    \caption{Upper panel: Non-thermal emissions from the SBN of M82 from hard X-ray to TeV gamma-ray domains from $\pi_0$ decay (dashed red line), inverse Compton scattering (dotted magenta line), bremsstrahlung radiation (dash-dotted green line), and synchrotron radiation (solid blue line). The flux is fitted to gamma-ray data from Fermi-LAT and VERITAS \citep{acero2015,acciari2009} and upper limits from Chandra \citep{strickland2007}. 
    Lower panel: Same as the upper panel but for Arp 220. Gamma-ray data from Fermi-LAT and VERITAS \citep{peng2016,fleischhack2015} and upper limits from Chandra \citep{paggi2017} are also overlaid. 
    }
    \label{fig:gamma_spectrum_M82_Arp220}
\end{figure}

The resulting fit of the non-thermal emissions is presented in Fig.~\ref{fig:gamma_spectrum_M82_Arp220} with the fit parameters as reported in Table \ref{tab:parameters}. We could also see that gamma-rays of energy above about $100$ MeV are mostly induced by CR protons. Below about $100$ MeV, leptonic processes namely bremsstrahlung radiation, inverse Compton scattering, and synchrotron radiation start to dominate the non-thermal emissions. As before, we adopt these fit parameters to predict the ionization rate for the SBN of M82 (see Fig. \ref{fig:ion}) which could reach $\zeta({\rm H}_2)\simeq 6.9\times 10^{-15}$ s$^{-1}$ for clouds with $N({\rm H}_2)=10^{23}$ cm$^{-2}$. The ionization rate in this case is about 2 times lower than that of NGC 253 even though the nucleus of M82 has a higher supernova rate. This is because the radius of the SBN of M82 is slightly larger and also it has a larger wind speed which results in particles escaping the nucleus more quickly via advection.

Another interesting system to be considered is the galaxy Arp 220 which is located at a distance of $d_{\rm gal}\simeq 77$ Mpc \citep{scoville1998}. In fact, this galaxy is a merger of two galaxies and, thus, it has two dense nuclei which are about a few hundred parsecs apart from each other. We shall follow \citet{peretti2019} and treat the Arp 220 SBN approximately as one nucleus with an effective size $R\simeq 165$ pc. Interestingly, the nucleus of this galaxy has been observed at a few different wavelengths in the radio domain and the spectral analysis of several detected sources allow us to infer the supernova rate of $\mathcal{R}_{\rm SNR}\simeq 2$-$6$ yr$^{-1}$. We fit the non-thermal spectrum of the Arp 220 nucleus to X-ray data from Chandra \citep{paggi2017} and gamma-ray data from Fermi-LAT and VERITAS \citep{peng2016,fleischhack2015}. The results are shown in Fig. \ref{fig:gamma_spectrum_M82_Arp220}. The underlying CR spectra leads to the CR-induced ionization rate of about $\zeta({\rm H}_2)\simeq 2.5\times 10^{-14}$ s$^{-1}$ for clouds with $N({\rm H}_2)=10^{23}$ cm$^{-2}$. The ionization rate versus column density is shown in Fig. \ref{fig:ion}. 

We summarize the ionization rate for clouds of column density $N({\rm H}_2)=10^{23}$ cm$^{-2}$ for all these prototypical SBNi in Table \ref{tab:ion}. The ionization rates from CR protons and electrons are also shown separately. It is interesting to note also that the contribution from CR protons is always slightly more dominant than that of electrons, especially at large column densities (see also the lower panel of Fig. \ref{fig:ion} for the case of NGC 253). Another point worth mentioning is that our predicted ionization rates for the three SBNi considered only differ slightly. 
These similarities, in fact, come from similarities in the fitted CR proton spectra. For the discussion of these similarities, let's parametrize for this discussion the CR proton spectrum as follows $j_{\rm p}(E)=j_{\rm p,300 \, GeV}g(E)/g(E=300\, {\rm GeV})$ where $g(E)$ is a function describing the shape of the spectrum and $j_{\rm p,300 \, GeV}$ is the normalization fixed at $E=300$ GeV. We should first notice that the CR transport is mostly dominated by energy loss (calorimetric systems) and, thus, the form of $g(E)$ is determined mostly by the injection spectral index of SNRs $\alpha$ (see Eq. \ref{eq:QSNRp}) which has rather similar values (between $4.2$ and $4.4$) as constrained by the gamma-ray spectral index. 
More importantly, since the gamma-ray fluxes are expected to be dominated by hadronic gamma rays, the normalization of the CR spectra should be, roughly speaking, proportional to the gamma-ray flux, distance squared, and the inverse of the total gas mass of the SBN (as also mentioned in the previous subsection). In fact, it can be shown that $j_{\rm p,300 \, GeV}\sim \phi_\gamma(E_\gamma=30\,{\rm GeV}) d^2/M_{\rm gas}\sim\phi_\gamma(E_\gamma=30\,{\rm GeV}) d^2/(n_{\rm ISM}R^3)$. As the values of $\phi_\gamma(E_\gamma=30\,{\rm GeV}) d^2/(n_{\rm ISM}R^3)$ are only different by a factor of a few for the three SBNi, the differences in the ionization rates are also expected to be of this order.

\begin{table*}
\centering
\caption{Parameters for both non-thermal and thermal emissions from the SBNi of NGC 253, M82, and Arp 220. We have fixed the distance $d_{\rm gal}$, the SBN size $R$, and the galactic wind speed $u_w$ as in \citet{peretti2019} as motivated by independent observations. The other parameters are fitted using data from radio to gamma-ray observations (see the comments for more details on parameters and their corresponding most constraining data). 
}
\label{tab:parameters}
\begin{tabular}{llcccc} 
\hline
Parameters & Description & Comments & NGC 253 & M82 & Arp 220\\
\hline
$d_{\rm gal}$ (Mpc) & Distance to the Milky Way & \multirow{4}{*}{\shortstack{Parameters being fixed as motivated by\\ independent observations or previous works}} & 3.8 & 3.9 & 77.0\\
$z_{\rm SBN}$ & Redshift of the SBN & & $8.8\times 10^{-4}$ & $9\times10^{-4}$ & $1.76\times 10^{-2}$\\
$R$ (pc) & Radius of the SBN & & 150 & 220 & 250\\
$u_{w}$ (km/s) & Speed of galactic wind & & 300 & 600 & 500\\
\hline
$\mathcal{R}_{\rm SNR}$ (yr$^{-1}$) & Supernova rate in the SBN & \multirow{3}{*}{\shortstack{Fitted parameters\\ mostly constrained by gamma-ray data}} & 0.03 & 0.05 & 2.25\\
$\alpha$ & Index of the CR injection spectrum & & 4.3 & 4.25 & 4.45\\
$n_{\rm ISM}$ (cm$^{-3}$) & ISM density in the SBN & & 170 & 155 & 3290\\
\hline
$B$ ($\mu$G) & Magnetic field strength & \multirow{3}{*}{\shortstack{Fitted parameters\\ mostly constrained by radio data}} & 120 & 150 & 500\\
$n_{\rm e}$ (cm$^{-3}$) & Density of thermal electrons in the SBN & & 30 & 22.75 & 87.5\\
$T_{\rm e}$ (K) & Temperature of thermal electrons in the SBN & & 8000 & 7000 & 3000\\
\hline
$U_{\rm FIR}$ (eV/cm$^3$) & Energy density of FIR photons & \multirow{8}{*}{\shortstack{Fitted parameters\\ constrained by data from far-infrared to optical}} & 979.0 & 455.0 & 15660.5\\
$k_{\rm B}T_{\rm FIR}$ (meV) & Temperature of FIR photons & & 3.5 & 3.0 & 3.5\\
$U_{\rm MIR}$ (eV/cm$^3$) & Energy density of MIR photons & & 293.5 & 318.5 & 4698.0\\
$k_{\rm B}T_{\rm MIR}$ (meV) & Temperature of MIR photons & & 8.75 & 7.5 & 7.0\\
$U_{\rm NIR}$ (eV/cm$^3$) & Energy density of NIR photons & & 293.5 & 227.5 & 62.5\\
$k_{\rm B}T_{\rm NIR}$ (meV) & Temperature of NIR photons & & 29.75 & 24.0 & 29.75\\
$U_{\rm OPT}$ (eV/cm$^3$) & Energy density of OPT photons & & 1468.0 & 273.0 & 783.0\\
$k_{\rm B}T_{\rm OPT}$ (meV) & Temperature of OPT photons & & 332.5 & 330.0 & 350.0\\
\hline
\end{tabular}
\end{table*}

\begin{table*}
\centering
\caption{Cosmic-ray induced ionization rates in SBN of NGC 253, M82, and Arp 220. Contribution to ionization rate from cosmic-ray protons, electrons, and total ionization rates are presented.}
\label{tab:ion}
\begin{tabular}{llccc} 
\hline
Ionization rate for $N({\rm H}_2)=10^{23}$ cm$^{-3}$ & Description & NGC 253 & M82 & Arp 220\\
\hline
$\zeta_{\rm p}({\rm H}_2)$ (s$^{-1}$) & Ionization rate from protons & $8.1\times 10^{-15}$ & $3.6\times 10^{-15}$ & $1.4\times 10^{-14}$\\
$\zeta_{\rm e}({\rm H}_2)$ (s$^{-1}$) & Ionization rate from electrons & $2.8\times 10^{-15}$ & $1.5\times 10^{-15}$ & $4.0\times 10^{-15}$ \\
$\zeta({\rm H}_2)=1.5\zeta_{\rm p}({\rm H}_2)+\zeta_{\rm e}({\rm H}_2)$ (s$^{-1}$) & Total ionization rate & $1.5\times 10^{-14}$ & $6.9\times 10^{-15}$ & $2.5\times 10^{-14}$\\
\hline
\end{tabular}
\end{table*}

\section{Discussions and Conclusions}
\label{sec:discussion}

We have performed fits for the non-thermal emissions to derive the CR spectra in the nuclei of some prototypical starburst galaxies namely NGC 253, M82, and Arp 220. These spectra are then implemented in the ballistic model \citep[see Section \ref{sec:ionization} and also][]{padovani2009} to describe the penetration of CRs into dense molecular clouds which allows us to predict the ionization rate for clouds of different column densities. Our predicted ionization rate varies around $10^{-14}$ s$^{-1}$ for all the prototypical SBNi considered and the values can decrease slightly with increasing column density.

In the case of NGC 253, our predicted ionization rate $\zeta({\rm H}_2)$ are compared to inferred values from molecular line observations by \citetalias{holdship2022} and \citetalias{behrens2022}. In most cases, the inferred ionization rates are a few times to about an order of magnitude higher than our predicted values. Such a discrepancy can be due to

(i) {\it Difference in the regions probed by the observations}: 
One important difference between our analysis and the ones from \citetalias{holdship2022} and \citetalias{behrens2022} is the size over which the modelling is performed. The inference of ionization rates by \citetalias{holdship2022} and \citetalias{behrens2022} focused in various molecular clouds of size comparable to the ALMA telescope synthetised beam of 1$\fs$6, that is roughly 30 pc. Our study, on the other hand, is based on observing constraints coming mostly from gamma-ray telescopes with large point-spread functions, typically of a few arcminutes (relatively low spatial resolution compared to radio observations of molecular line emissions). This requires us to assume a uniform CR density over the entire SBN in our modeling which should be a good approximation given the high supernova rate within
the system. Variations of CR density on small scales, however, might exist on scales comparable to remnant size \cite{phan2021,phan2023} which should lead to corresponding variations on ionization rates. Modelling such variations might require a better description of not only CR transport but also of the large-scale ISM within these SBNi (relevant for energy loss processes of CRs) and will be examined in our future works.

(ii) {\it Uncertainties in chemical modeling of line observations}: The two analyses, \citetalias{holdship2022} and \citetalias{behrens2022}, examine the same set of clouds using different line emissions; \citetalias{holdship2022} study H$_3$O$^+$ and SO and \citetalias{behrens2022} focus on HCN and HNC. The results, however, contain values of ionization rates different by up to an order of magnitude which could mean that the uncertainties in these inferred values are not fully reflected by the errors on ionization rates. In this scenario, it would be interesting to perform a combined analysis taking into account not only line emission data but also non-thermal emission data. Such an analysis might help to reduce the uncertainties on the inferred ionization rates. 

(iii) {\it Uncertainties in gas mass and supernova rate}: As mentioned in Section \ref{sec:SBNi}, the normalization CR proton spectrum as constrained by gamma-ray data should be, roughly speaking, proportional to the gamma-ray flux and the inverse of the SBN gas mass, i.e. $j_{\rm p,300 \,GeV}\sim \phi_\gamma(E_\gamma=30\,{\rm GeV})/M_{\rm gas}$. In this work, the SBN size and the fitted ISM density correspond to $M_{\rm gas}\simeq 7\times 10^7 M_\odot$ which is actually within the range of the gas mass estimated independently from molecular line observations (between $2.5\times 10^7M_\odot$ and $4\times 10^8 M_\odot$ as presented respectively by \citealt{harrison1999} and \citealt{houghton1997}, see also the discussion by \citealt{bradford2003}).
However, the uncertainty on the gas mass estimate might also mean that its value is actually a few times smaller than expected \citep[see e.g.][]{mauersberger1996} which should accordingly require a larger density of CR protons in order to fit the gamma-ray data and, as a result, lead to predicted ionization rates being higher. This can help to improve the agreement between our predictions and the measurements from \citetalias{holdship2022} and \citetalias{behrens2022}. Similarly, the uncertainty on the gamma-ray flux, which can be translated into the uncertainty on the fitted value of the supernova rate (see discussions in Section \ref{sec:SBNi}), can also be a source of the discrepancy. Indeed, there exists also the possibility that the SBN gas mass is close to the higher end of its uncertainty range and the supernova rate is lower than expected which could further increase the difference between our predictions and measurements. More precise estimates of these quantities are, therefore, essential to improve our understanding of this discrepancy.

(iv) {\it Local sources of MeV CRs}: For the SBNi considered, the non-thermal emissions in the GeV and TeV energy, where data are most constraining, are contributed mostly from the decay of $\pi_0$ created in proton-proton interactions. The production of $\pi_0$, however, has a threshold $E_{\rm th}\simeq 280$ MeV meaning that these gamma-ray data could not probe CR protons with $E\lesssim E_{\rm th}$. In other words, there might exist a class of sources accelerating mostly CRs in the energy range of around a few hundred MeVs, e.g. wind termination shocks of stars \citep{scherer2008}, protostellar jets embedded within molecular clouds \citep{padovani2015,padovani2016,gaches2018}, or even H\rom{2} regions \citep{padovani2019,meng2019}, that contribute to the ionization rate in these systems but could not be observed with GeV and TeV gamma-ray telescopes. We note also that if these MeV sources exist in SBNi and they are sufficiently abundant, they might contribute to the gamma-ray emissions in the MeV energy range, particularly relevant for future missions like {\rm eASTROGRAM} or {\rm AMEGO} \citep{deangelis2018,mcenery2019}.  

Further investigations are required to understand the discrepancies between our predicted ionization rates and the values inferred from molecular observations. The difference by a factor of a few in most cases, however, mean that our predictions for ionization rates in SBNi can be potentially very useful for future chemical and dynamical studies of these rather complex star-forming regions.    

\section*{Acknowledgements}
VHMP acknowledges support from the Initiative Physique des Infinis (IPI), a research training program of the Idex SUPER at Sorbonne Universit\'e. The research activity of EP was supported by Villum Fonden (project No.~18994) and by the European Union’s Horizon 2020 research and innovation program under the Marie Sklodowska-Curie grant agreement No. 847523 "INTERACTIONS". EP was also supported by Agence Nationale de la Recherche (grant ANR-21-CE31-0028). We thank Stefano Gabici, Dieter Breitschwerdt, and Erica Behrens for fruitful discussions and valuable comments. 




\bibliographystyle{mnras}
\bibliography{mybib}



\appendix



\section{Emissions in the radio, infrared and optical domain}
\label{appendixA}
We present in this appendix comparisons between the fitted spectra of thermal and non-thermal emissions and data from various observations retrieved from the {\it NASA/IPAC Extragalactic Database} in the domain from radio to optical frequency for the three SBNi of NGC 253, M82, and Arp 220. 

Note that we have taken into account the free-free absorption in the radio domain. The opacity for this process can be evaluated as follows (see Chapter 5 of \citealt{rybicki1986})
\begin{eqnarray}
    && \tau_{\rm ff}(E_\gamma)\simeq 3.7\times10^8 \left(\frac{R}{1\,{\rm cm}}\right)\bar{g}_{\rm ff}\left(\frac{T_{\rm e}}{1\,{\rm K}}\right)^{-1/2}\left(\frac{n_{\rm e}}{1\,{cm}^{-3}}\right)^2\n\\
    &&\qquad\qquad\times\left(\frac{E_\gamma/h}{1\,{\rm Hz}}\right)^{-3}\left[1-\exp\left(\frac{E_\gamma}{k_{\rm B}T_{\rm e}}\right)\right],
\end{eqnarray}
where $\bar{g}_{\rm ff}$ is the Gaunt factor 
\begin{eqnarray}
    \bar{g}_{\rm ff}\simeq 
    {\rm min}\left\{1,\frac{\sqrt{3}}{\pi}\ln\left[0.945\left(\frac{k_{\rm B}T_{\rm e}}{E_\gamma}\right)\left(\frac{k_{\rm B}T_{\rm e}}{13.6\,{\rm eV}}\right)^{1/2}\right]\right\}.
\end{eqnarray}

\begin{figure}
	\includegraphics[width=3.5in]{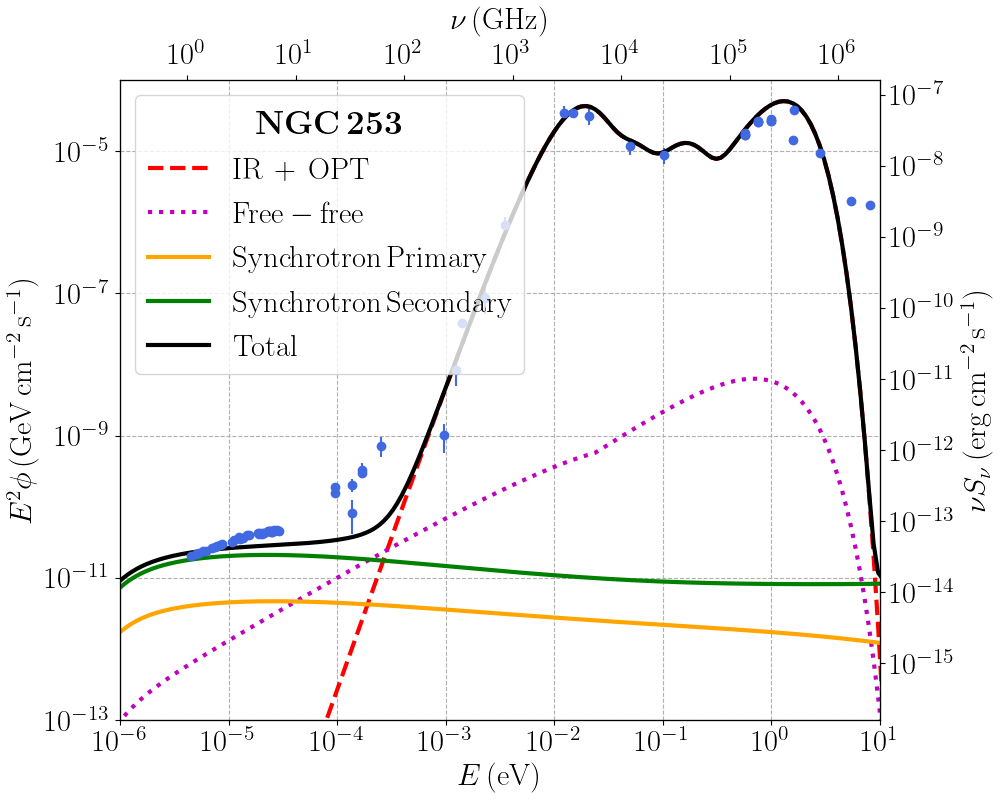}\\
	\includegraphics[width=3.5in]{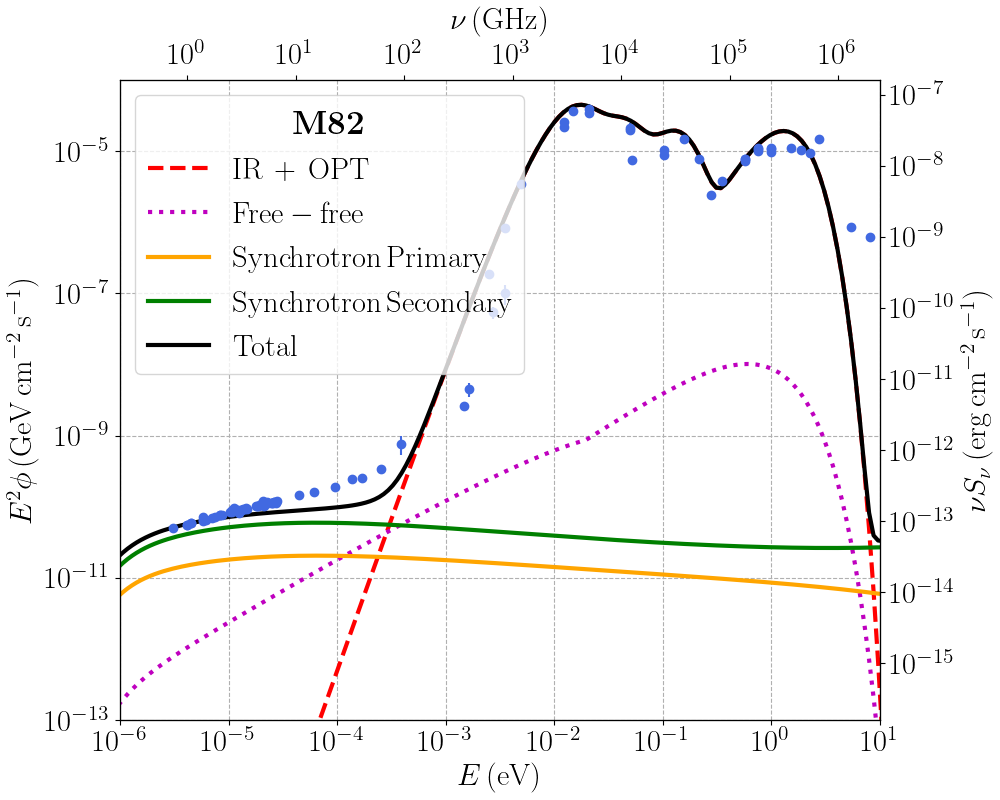}\\
	\includegraphics[width=3.5in]{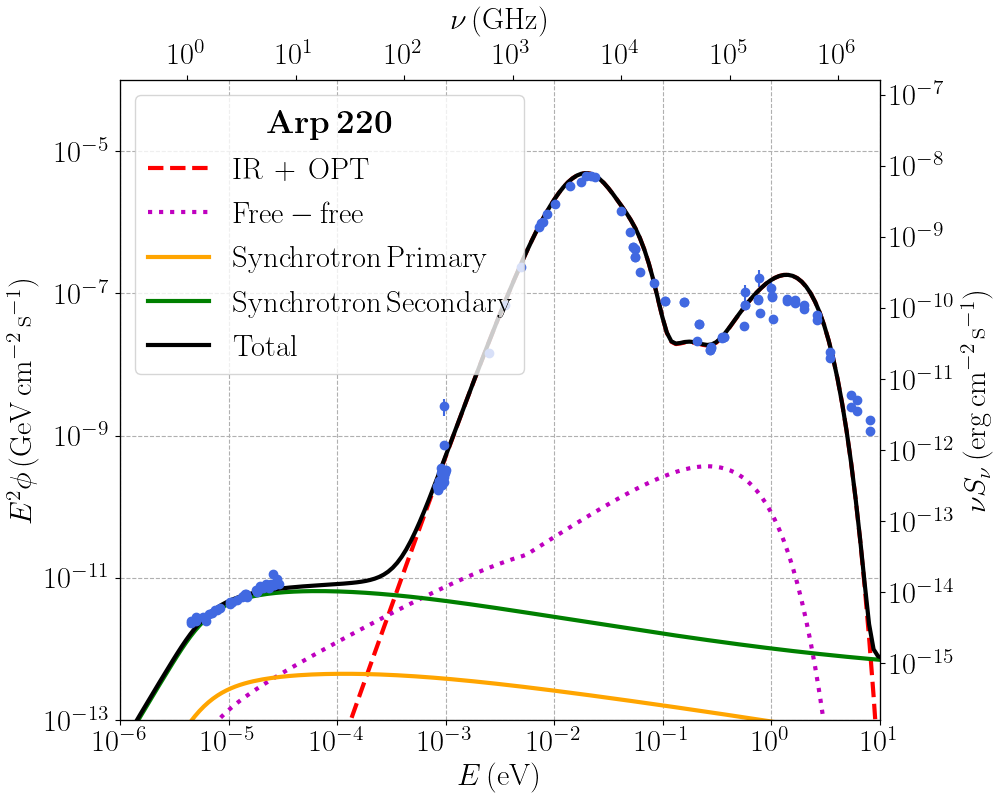}
    \caption{Fitted spectra in the domain from radio to optical frequency for the three starburst nuclei of NGC 253, M82, and Arp 220 overlaid with data from various observations retrieved from the {\it NASA/IPAC Extragalactic Database}.}
    \label{fig:radio}
\end{figure}


\bsp	
\label{lastpage}
\end{document}